
\input phyzzx
\singlespace
\twelvepoint
%
%
\font\mathbold=cmmib10 at12pt
\font\bf=cmbx10 at12pt
\newfam\bffam \def\mbf{\fam\bffam\mathbold} \textfont\bffam=\mathbold
\mathchardef\beta="710C
\mathchardef\gamma="710D
\mathchardef\eta="7111
\mathchardef\xi="7118
\def\mbB{{\mbf B}}
\def\mbC{{\mbf C}}
\def\mbbeta{{\mbf \beta}}
\def\mbgamma{{\mbf \gamma}}
\def\mbeta{{\mbf \eta}}
\def\mbxi{{\mbf \xi}}
\def\mbT{{\mbf T}}
\def\mbW{{\mbf W}}
\def\mbV{{\mbf V}}
%
\REF\bv{N. Berkovits and C. Vafa, {\sl Mod. Phys. Lett.} {\bf A9} 653
(1994).}
\REF\kun{H. Kunitomo, {\sl On the Nonlinear Realization of the
Superconformal
Symmetry}, preprint OU-HET 196, hep-th/9407052 (1994).}
\REF\bop{F. Bastianelli, N. Ohta and J.L. Petersen, {\sl A Hierarchy
of
Superstrings},
                 preprint NBI-HE-94-20, hep-th/9403150 (1994).}
\REF\bfw{N. Berkovits, M. Freeman and P. West, {\sl Phys. Lett.} {\bf
B325} 63
(1994).}
\REF\kst{H. Kunitomo, M. Sakaguchi and A. Tokura, {\sl A Universal
$w$\ String
Theory}, preprint OU-HET 187, hep-th/9403086 (1994), to be published
in {\sl
Int. J. Mod. Phys.} {\bf A}.}
\REF\li{K. Li, {\sl Linear $w_N$\ Gravity}, preprint CALT-68-1724
(1991).}
\REF\ik{H. Ishikawa and M. Kato, {\sl Mod. Phys. Lett.} {\bf A9} 725
(1994).}
\REF\bo{N. Berkovits and N. Ohta, {\sl Embeddings for Non-Critical
Superstrings}, preprint KCL-TH-94-6, OU-HET 189, hep-th/9405144
(1994).}
\REF\ko{T. Kugo and I. Ojima, {\sl Porg. Theor. Phys. Supplement}
{\bf 66}
(1979) 1.}
\REF\b{F. Bastianelli, {\sl Phys. Lett.} {\bf 322B} (1994) 340.}
\REF\t{K. Thielemans, {\sl Int. J. Mod. Phys.} {\bf C2} (1991) 787.}
\REF\tk{K. Thielemans and S. Krivonos, unpublished.}
%
\pubnum={OU-HET 199}
\titlepage
\title{A Hierarchy of Super $w$\ Strings}
\author{Hiroshi KUNITOMO\foot{e-mail: kunitomo@oskth.kek.jp},
Makoto SAKAGUCHI\foot{e-mail: gu@oskth.kek.jp} \break
and\break
Akira TOKURA\foot{e-mail: atoku@oskth.kek.jp}}
\address{\null\hskip-8mm
Department of Physics, Osaka University, \break
Toyonaka, Osaka 560, JAPAN}
\abstract{It is shown that the N-extended super $w_n$\ string can be
obtained
as a special class of vacua of both the N-extended super $w_{n+1}$\
string and
the (N+1)-extended super $w_n$\ string.
This hierarchy of super $w$\ string theories includes as
subhierarchies both
the superstring hierarchy given by Bastianelli, Ohta and Petersen and
the $w$\
string hierarchy given by the authors.
}
\endpage
\chapter{Introduction}

The bosonic string theory can be regarded as a reparametrization
invariant two
dimensional field theory. We can similarly consider an extended
string theory
when the two dimensional field theory has larger local symmetry.
Superstring
theory is such an example that is invariant under the local
supersymmetry.
In this way, there are several string theories corresponding to their
local
symmetries. After fixing these gauge invariances to the (generalized)
conformal
gauge, these two dimensional field theories become conformal field
theory with
extended conformal symmetries. Therefore a string theory can be
considered for
each extension of the conformal algebra.

In the above sense, it has been found that a string theory can be
viewed as a
class of vacua for the other string theory, where the latter theory
is based on
a larger extended conformal algebra including the one of the former
theory as a
subalgebra. The first example of such a realization was studied by
Berkovits
and Vafa\refmark{\bv}. They showed that the bosonic ($N=1$) string
theory can
be viewed as a particular class of vacua for the $N=1$\ ($N=2$)
superstring,
which can be explained by means of a spontaneous symmetry breaking of
superconformal symmetry\refmark{\kun}. The larger superconformal
symmetry is
realized by introducing Nambu-Goldstone (NG) fields transformed
inhomogeneously
under the broken symmetry. This realization was generalized to
general
$N$-extended superstrings by Bastianelli, Ohta and
Petersen\refmark{\bop}. They
found that the $N$-extended superstring can be understood as a
spontaneously
broken phase of the $(N+1)$-extended superstring and therefore there
is a
hierarchy of superstrings related by symmetry breaking.

It was also found that there are similar realizations for $W$\ string
theories\refmark{\bfw,\kst}. The bosonic string can be realized as a
broken
phase of the nonlinear $W_3$\ string\refmark{\bfw} and it can be
generalized to
general ($N=1$) linear $W_n$\ strings by using the linear $W_n$\
algebra
introduced in Ref. [\li]\refmark{\kst}. We denote this linear $W_n$\
string
simply by $w_n$\ string using lowercase $w$. This hierarchy of $w$\
strings has
a different structure from the one of the superstrings since the
$w_n$\ algebra
is not a subalgebra of the $w_{n+1}$\ algebra. While the hierarchy
relate the
$w_n$\ string to the $w_{n+1}$\ string ($n>2$), both theories are in
the broken
phase. Only the bosonic ($w_2$) string in this hierarchy can be in
the general
phase\refmark{\kst}.

In this paper, we construct a hierarchy of super $w$\ strings which
is a
natural generalization of the above two hierarchies. We show that
general
$N$-extended super (N-super) $w_n$\ string can be understood as a
broken phase
of both $N$-super $w_{n+1}$\ string and $(N+1)$-super $w_n$\ string.
The
hierarchy is parametrized by two integers $(N,n)$\ and includes the
superstring
hierarchy $(N,2)$\ and the $w$\ string hierarchy $(0,n)$\ as
subhierarchies.

This paper is organized as follows.

Before constructing the super $w$\ string hierarchy, we give an
algebraic
definition of the $N$-super $w_n$\ string in Sec. 2. We introduce
$N$-extended
superspace ($N$-superspace) and write the $N$-super $w_n$\ algebra by
using the
$N$-extended superfields ($N$-superfields). Introducing the ghost
$N$-superfields, the BRST charge is given by a $N$-superspace
integral. For
complete definition of the string theory, we must also give rules for
calculating amplitudes which are not given in this paper. They are
discussed in
the final section. We also give the $(N+1)$-super $w_n$\ string by
using the
$N$-superspace. This is useful for constructing the $N$-super $w_n$\
string as
a broken phase of the $(N+1)$-super $w_n$\ string explained in Sec.
4.

In Sec. 3, we construct such a realization of the $N$-super
$w_{n+1}$\ string
as is equivalent to the $N$-super $w_n$\ string. We first explain the
results
for the $N=0$\ case obtained in the Ref. [\kst]. By introducing NG
fields for
each broken generator, the generators of the $w_{n+1}$\ algebra are
constructed. The equivalence of cohomology is shown by finding the
similarity
transformation which maps the BRST charge of the $w_{n+1}$\ string to
the sum
of the one of the $w_n$\ string and a topological BRST charge. Since
the
cohomology of the topological BRST charge is trivial, the cohomology
of two
string theories coincide\refmark{\ik}. Then these results are
extended to the
general $N$-super $w$\ strings by using the $N$-superspace. This is a
generalization of the $w$\ string hierarchy obtained in Ref. [\kst].

The $N$-super $w_n$\ string can also be obtained as a broken phase of
the
$(N+1)$-super $w_n$\ string, which is a generalization of the
superstring
hierarchy given in Ref. [\bop]. We consider it in Sec. 4 in a similar
way to
Sec. 3. The realization of the $(N+1)$-super $w_n$\ algebra is
constructed by
using the NG fields. The similarity transformation is found by
separating it to
three parts and guarantees the equivalence of the cohomology. By
combining the
results obtained in Sec. 3, we have a hierarchy of super $w$\ string
parametrized two integers $(N,n)$. The $N$-super $w_n$\ string can be
understood as a broken phase of both the $N$-super $w_{n+1}$\ string
and the
$(N+1)$-super $w_n$\ string. This super $w$\ string hierarchy is
including both
the superstring hierarchy\refmark{\bo} $(N,2)$\ and the $w$\ string
hierarchy\refmark{\kst} $(0,n)$\ as subhierarchies.

The Sec. 5 is devoted to discussion.

\chapter{$N$-super $w_n$\ string}

Before studying the super $w$\ string hierarchy, we give an algebraic
definition of the $N$-super $w_n$\ string by constructing the BRST
charge which
determine physical states of the theory. We do not mention the rules
of
calculating amplitudes although they are necessary to the complete
definition.
We discuss it in the final section.

Operator product expansion (OPE) of the linear $w_n$\ algebra
introduced in
Ref. [\li] has the form
$$
\eqalignno{
W_i(z)W_j(w)&\sim {c(0,n)\delta_{i,0}\delta_{i,j}\over 2(z-w)^4}+
{(i+j+2)W_{i+j}(w)\over (z-w)^2}+{(i+1)\partial W_{i+j}(w)\over
z-w},\cr
&&{\rm for } \ i+j\le n-2\cr
W_i(z)W_j(w)&\sim 0,\qquad\qquad{\rm for }\  i+j>n-2
&\eqname\w\cr
}
$$
where generators $\{W_i(z);i=0,\cdots n-2\}$\ have dimensions $i+2$\
and
$W_0(z)=T(z)$\ is the stress tensor. When the central charge is equal
to the
critical value $c(0,n)=2(n-1)(2n^2+2n+1)$, we can define the $w_n$\
string
theory by introducing fermionic ghost fields $(B_i(z),C_i(z))\
(i=0,\cdots ,
n-2)$\ with dimensions $(i+2,-i-1)$. The nilpotent BRST charge is
defined by
$$
Q_{BRST}(0,n)=\oint {dz\over 2\pi i}
\sum_{i=0}^{n-2} C_i(W_i+{1\over 2}W_i^{gh}),
$$
where
$$
W_i^{gh}=\sum_{j=0}^{n-i-2}\left(-(i+j+2)B_{i+j}\partial C_j
-(j+2)\partial B_{i+j}C_j\right).
$$
We extend this $w_n$\ string theory to the $N$-super ($N\ge 1$)
$w_n$\ string
theory.

We use the $N$-superspace to describe the $N$-super $w_n$\ algebra.
The
coordinates of the $N$-superspace are denoted by $Z=(z,\theta^a)$\
with
$a=1,\cdots N$. The other notations and conventions related to the
superspace
are\refmark{\bop}
$$
\eqalignno{
D^a=&{\partial\over\partial\theta^a}+\theta^a\partial,\qquad
\qquad\qquad\quad
\{D^a,D^b\}=2\delta^{a,b}\partial ,\cr
Z_{12}=&z_1-z_2-\theta_1^a\theta_2^a,\qquad\qquad\quad
\Theta_{12}^a=\theta_1^a-\theta_2^a,\cr
\Theta_{12}^N=&{1\over N!}\epsilon^{a_1\cdots
a_N}\Theta_{12}^{a_1}\cdots\Theta_{12}^{a_N},
\qquad
(\Theta_{12}^{N-1})^a=
{1\over (N-1)!}\epsilon^{b_1\cdots
b_{N-1}a}\Theta_{12}^{b_1}\cdots\Theta_{12}^{b_{N-1}},\cr
\oint dZ=&\oint{dz\over 2\pi i}\int d\theta^N,\qquad\qquad\quad
d\theta^N={1\over N!}\epsilon^{a_1\cdots a_N}d\theta^{a_N}\cdots
d\theta^{a_1}.
}
$$

The $N$-super $w_n$\ algebra is generated by supercurrents
$\{\mbW_i(Z);\
i=0,\cdots n-2\}$\ which have dimensions $i+2-N/2$\ and
$\mbW_0(Z)=\mbT(Z)$\ is
the superstress tensor. The OPE defining the $N$-super $w_n$\ algebra
is
$$
\eqalignno{
\mbW_i(Z_1)\mbW_j(Z_2)\sim&
(-1)^{N-1}{\hat c(N,n)\delta_{i,0}\delta_{i,j}\over 2Z_{12}^{4-N}}+
{\Theta_{12}^N\over Z_{12}^2}(i+j+2-{N\over 2})\mbW_{i+j}(Z_2)\cr
&+{\Theta_{12}^N\over Z_{12}}(i+1)\partial\mbW_{i+j}(Z_2)+
{(\Theta_{12}^{N-1})^a\over Z_{12}}{1\over
2}D^a\mbW_{i+j}(Z_{12}),\cr
&&{\rm for } \ i+j\le n-2\cr
\mbW_i(Z_1)\mbW_j(Z_2)\sim& 0, \qquad\qquad{\rm for }\  i+j>n-2
&\eqname\Nsuperw\cr
}
$$
where the central extension term must vanish for $N\ge 4$. The
critical central
charge is obtained by calculating central charge of corresponding
ghost fields
and given by
$$
\eqalignno{
\hat c(N,n)=&{(3-N)!\over 3!}c(N,n)\cr
=&\cases {(n-1)(2n+1) & for $N=1$\cr
	  (n-1) & for $N=2$\cr
	   0 & for $N\ge 3$.}
&\eqname\critical\cr
}
$$
Here we note that the critical central charge for $N\ge 3$\ is zero,
which is
consistent to the above argument of the central extension.

When the central charge is equal to this critical value, we can
construct the
$N$-super $w_n$\ string theory by introducing ghost fields
$(\mbB_i(Z),
\mbC_i(Z))\ (i=0,\cdots ,n-2)$\ satisfying
$$
\mbC_i(Z_1)\mbB_j(Z_2)\sim {\Theta_{12}^N\over Z_{12}}\delta_{i,j}.
$$
The ghost superfields $\mbC_i(Z)$\ are fermionic with dimension
$(-i-1)$ and
the anti-ghost superfields $\mbB_i(Z)$\ are fermionic (bosonic) for
$N=$\ even
(odd) with dimension $(i+2-N/2)$.
The nilpotent BRST charge is given by the $N$-superspace integral as
$$
\eqalignno{
Q_{BRST}(N,n)=&\oint dZ \sum_{i=0}^{n-2}\mbC_i(\mbW_i+{1\over
2}\mbW_i^{gh}),
&\eqname\BRS\cr
}
$$
where
$$
\mbW_i^{gh}(Z)=\sum_{j=0}^{n-i-2}\left(-(i+j+2-{N\over
2})\mbB_{i+j}\partial
\mbC_j
-(j+1)\partial \mbB_{i+j}\mbC_j-{(-1)^N\over
2}D^a\mbB_{i+j}D^a\mbC_j\right).
$$
The cohomology of this BRST charge describes physical states of the
theory.
Therefore we have defined, at least free, string theory.
For calculating amplitudes, we must further study global degrees of
freedom,
which is discussed in the final section.

Before closing this section, we write the $(N+1)$-super $w_n$\ string
theory by
using the $N$-superspace for later convenience. We restrict the
central charge
to the critical value \critical\ in the remaining part of this
section.

We consider first the case of $N=0$. The $N=1\ w_n$\ algebra is
generated by
two types of ordinary currents $(W_i(z),V_i(z))\ (i=0,\cdots ,n-2)$\
with
dimensions $(i+2,i+3/2)$\ as\refmark{\kst}
$$
\eqalignno{
W_i(z)W_j(w)\sim&{3(n-1)(2n+1)\delta_{i,0}\delta_{i,j}\over 2(z-w)^4}
+{(i+j+2)W_{i+j}(w)\over (z-w)^2}
+{(i+1)\partial W_{i+j}(w)\over z-w},\cr
W_i(z)V_j(w)\sim&{(i+j+{3\over 2})V_{i+j}(w)\over (z-w)^2}
+{(i+1)\partial V_{i+j}(w)\over z-w},\cr
V_i(z)V_j(w)\sim&{2(n-1)(2n+1)\delta_{i,0}\delta_{i,j}\over
(z-w)^3}+{2W_{i+j}(w)\over z-w},
\qquad\qquad {\rm for}\ i+j\le n-2 \cr
W_i(z)W_j(w)\sim&W_i(z)V_j(w)\sim V_i(z)V_j(w)\sim 0.
\qquad\qquad {\rm for}\ i+j>n-2
}
$$
We have to introduce both fermionic ghosts $(B_i,C_i)$\ and bosonic
ghosts
$(\beta_i,\gamma_i)$\ $(i=0,\cdots,n-2)$\ with dimensions
$(i+2,-i-1)$\ and
$(i+3/2,-i-1/2)$. The BRST charge $Q_{BRST}(1,n)$\ is rewritten
as\foot{
We are taking a different convention from Ref[\kst] which are
obtained by
$V_i^{(gh)}\rightarrow -V_i^{(gh)}$.}
$$
Q_{BRST}(1,n)=\oint {dz\over 2\pi
i}\sum_{i=0}^{n-2}\Big(C_i(W_i+{1\over
2}W_i^{gh})
+\gamma_i(V_i+{1\over 2}V_i^{gh})\Big),
$$
where
$$
\eqalignno{
W_i^{gh}=&W_i(BC)+W_i(\beta\gamma) ,\cr
=&\sum_{j=0}^{n-i-2}\Big(-(i+j+2)B_{i+j}\partial C_j-(j+1)\partial
B_{i+j}C_j\Big)\cr
&+\sum_{j=0}^{n-i-2}\Big(-(i+j+{3\over
2})\beta_{i+j}\partial\gamma_j-(j+{1\over
2})\partial\beta_{i+j}\gamma\Big),\cr
V_i^{gh}=&\sum_{j=0}^{n-i-2}\Big(-2B_{i+j}\gamma_j
+(i+j+{3\over 2})\beta_{i+j}\partial
C_j+(j+1)\partial\beta_{i+j}C_j\Big).
}
$$

For $N\ge 1$, the $(N+1)$-super $w_n$\ algebra is rewritten by using
two
$N$-supercurrents $(\mbW_i(Z),\mbV_i(Z))$\ $(i=0,\cdots ,n-2)$\ with
dimensions
$(i+2-N/2,i+3/2-N/2)$. The OPE can be written by using the
$N$-superspace as
$$
\eqalignno{
\mbW_i(Z_1)\mbW_j(Z_2)\sim&{(n-1)\delta_{N,1}\delta_{i,j}\delta_{i,0}
\over
Z_{12}^3}
+{\Theta_{12}^N\over Z_{12}^2}(i+j+2-{N\over 2})\mbW_{i+j}(Z_2)\cr
&+{\Theta_{12}^N\over Z_{12}}(i+1)\partial \mbW_{i+j}(Z_2)
+{(\Theta_{12}^{N-1})^a\over Z_{12}}{1\over 2}D^a\mbW_{i+j}(Z_2),\cr
\mbW_i(Z_1)\mbV_j(Z_2)\sim&
{\Theta_{12}^N\over Z_{12}^2}(i+j+{3\over 2}-{N\over
2})\mbV_{i+j}(Z_2)\cr
&+{\Theta_{12}^N\over Z_{12}}(i+1)\partial\mbV_{i+j}(Z_2)
+{(\Theta_{12}^{N-1})^a\over Z_{12}}{1\over 2}D^a\mbV_{i+j}(Z_2),\cr
\mbV_i(Z_1)\mbV_j(Z_2)\sim&
-{2(n-1)\delta_{N,1}\delta_{i,0}\delta{i,j}\over Z_{12}^2}
-{\Theta_{12}^N\over Z_{12}}2\mbW_{i+j}(Z_2),
\qquad\qquad {\rm for}\ i+j\le n-2 \cr
\mbW_i(Z_1)\mbW_j(Z_2)\sim&\mbW_i(Z_1)\mbV_j(Z_2)\sim
\mbV_i(Z_1)\mbV_j(Z_2)\sim 0.
\qquad\qquad {\rm for}\ i+j>n-2
&\eqname\NoneN\cr
}
$$
The ghost superfields are $(\mbB_i,\mbC_i)$\ and
$(\mbbeta_i,\mbgamma_i)$\
$(i=0,\cdots,n-2)$\ with dimensions $(i+2-N/2,-i-1)$\ and
$(i+3/2-N/2,-i-1/2)$,
where ghost superfields $\mbC_i$\ are always fermionic and
$\mbgamma_i$\ are
always bosonic. On the other hand, anti-ghost superfields $\mbB_i\
(\mbbeta_i)$\ are fermionic (bosonic) for $N=$\ even and bosonic
(fermionic)
for $N=$\ odd. The BRST charge $Q_{BRST}(N,n)$\ is obtained by the
$N$-superspace integral as
$$
\eqalignno{
Q_{BRST}(N+1,n)=&\oint dZ \sum_{i=0}^{n-2}\Big(\mbC_i(\mbW_i+{1\over
2}\mbW_i^{gh})
+\mbgamma_i(\mbV_i+{1\over 2}\mbV_i^{gh})\Big),
}
$$
where
$$
\eqalignno{
\mbW_i^{gh}=&\mbW_i(\mbB\mbC)+\mbW_i(\mbbeta\mbgamma) ,\cr
=&\sum_{j=0}^{n-i-2}\Big(-(i+j+2-{N\over 2})\mbB_{i+j}\partial\mbC_j
-(j+1)\partial\mbB_{i+j}\mbC_j-{(-1)^N\over
2}D^a\mbB_{i+j}D^a\mbC_j\Big) \cr
&+(-1)^N\sum_{j=0}^{n-i-2}\Big(-(i+j+{3\over 2}-{N\over
2})\mbbeta_{i+j}\partial\mbgamma_j-(j+{1\over
2})\partial\mbbeta_{i+j}\mbgamma_j
+{(-1)^N\over 2}D^a\mbbeta_{i+j}D^a\mbgamma_j\Big),\cr
\mbV_i^{gh}=&\sum_{j=0}^{n-i-2}\Big(2\mbB_{i+j}\mbgamma_j
+(i+j+{3\over 2}-{N\over 2})\mbbeta_{i+j}\partial\mbC_j
+(j+1)\partial\mbbeta_{i+j}\mbC_j
-{(-1)^N\over 2}D^a\mbbeta_{i+j}D^a\mbC_j\Big).
}
$$

\chapter{The $N$-super $w_n$\ string as a broken phase of  the
$N$-super
$w_{n+1}$\ string}

In this section we generalize the $w$\ string hierarchy\refmark{\kst}
to the
$N$-super $w$\ string hierarchy with a fixed $N$. The $N$-super
$w_n$\ string
can be obtained as a broken phase of the $N$-super $w_{n+1}$\ string.

For making this paper self-contained, we first explain the results on
the
$(N=0)\ w$\ string hierarchy obtained in Ref. [\kst]. The $N=0\
w_{n+1}$\
algebra can be realized by means of a general stress tensor
$T^{(m)}(z)$\ with
$c=26$\ (unbroken generator) and the bosonic NG fields
$(\beta_i(z),{\gamma}_i(z))\ (i=1,\cdots ,n-1)$\ with dimensions
$(i+2,-i-1)$.
The generators which satisfy the OPE \w\ of the $w_{n+1}$\ algebra
are written
as
$$
W_i(z)=\delta_{i,0} T^{(m)}+i{\beta}_i+
\sum_{j=0}^{n-i-1}\Big(-(i+j+2){\beta}_{i+j}\partial{\gamma}_j
-(j+1)\partial{\beta}_{i+j}{\gamma}_j\Big).
$$
We can find a similarity transformation by which the BRST charge
\BRS\
$Q_{BRST}(0,n+1)$\ is transformed into the sum of the BRST charges of
the
$w_n$\ string $Q_{BRST}(0,n)$\ and a topological BRST charge $Q_t$:
$$
\eqalignno{
&e^RQ_{BRST}(0,n+1)e^{-R}=Q_{BRST}(0,n)+Q_t,\qquad\qquad
Q_t=\oint {dz\over 2\pi i}(n-1)C_{n-1}{\beta}_{n-1},\cr
&R={1\over n-1}\oint {dz\over 2\pi i} \sum_{i=0}^{n-2}
C_i\Big((n+1)B_{n-1}\partial{\gamma}_{n-i-1}+(n-i)\partial
B_{n-1}{\gamma}_{n-i-1}\Big).
}
$$
Since the cohomology of the topological BRST charge $Q_t$\ is
trivial, the
cohomology of $Q_{BRST}(0,n+1)$ coincides with that of
$Q_{BRST}(0,n)$. The
physical spectrum of both theories are identical. Here we should note
that the
above BRST charge $Q_{BRST}(0,n)$\ (or $w_n$\ algebra) is still
written in
terms of the NG fields $(\beta_i(z),{\gamma}_i(z))\ (i=1,\cdots
,n-2)$. Thus
the $w_n$\ string theory obtained in this way is in the broken phase.
This is
due to the fact that the $w_n$\ algebra is not a subalgebra of the
$w_{n+1}$\
algebra. Only the $w_2$\ (bosonic) string can be in the unbroken
phase. We
should also note that, for completing the proof of the equivalence,
we must
further show the cancellation of zero modes since the string
amplitudes do not
consist of only BRST invariant quantities\refmark{\bv,\kst}. The BRST
invariance holds only after integrations of the moduli.

Employing the $N$-superspace, we can easily generalize the above
construction
to the $N$-super $w$\ string hierarchy. The $N$-super $w_{n+1}$\
algebra can be
realized by the general superstress tensor $T^{(m)}(Z)$\ with $c=15,\
6\ {\rm
and}\ 0$\ for $N=1,\ 2\ {\rm and}\ \ge 3$\ and the NG superfields
$(\mbbeta_i(Z),\mbgamma_(Z))\ (i=1,\cdots ,n-1)$\ with dimensions
$(i+2-N/2,-i-1)$. $N$-super $w_{n+1}$\ algebra is realized as
$$
\eqalignno{
\mbW_i=&\delta_{i,o}\mbT^{(m)}+i\mbbeta_i\cr
&+\sum_{j=1}^{n-i-1}\Big((-1)^{N+1}(i+j+2-{N\over
2})\mbbeta_{i+j}\partial\mbgamma_j
+(-1)^{N+1}(j+1)\partial\mbbeta_{i+j}\mbgamma_j+{1\over
2}D^a\mbbeta_{i+j}D^a\mbgamma_j\Big),\cr
&&\eqname\superwre\cr
}
$$
where ${\beta}_i$\ are bosonic (fermionic) for $N=$\ even (odd) and
${\gamma}_i$\ are always bosonic superfields. The BRST charge
$Q_{BRST}(N,n+1)$\ is transformed into the sum of $Q_{BRST}(N,n)$\
and a
topological BRST charge $Q_t$\ by the similarity transformation as
$$
\eqalignno{
&e^RQ_{BRST}(N,n+1)e^{-R}=Q_{BRST}(N,n)+Q_t,\qquad\qquad
Q_t=(n-1)\oint dZ\mbC_{n-1}\mbbeta_{n-1},\cr
&R={1\over n-1}\oint dZ\sum_{i=0}^{n-2}
\mbC_i\Big((n+1-{N\over
2})\mbB_{n-1}\partial\mbgamma_{n-i-1}+(n-i)\partial
\mbB_{n-1}\mbgamma_{n-i-1}\cr
&\qquad\qquad\qquad\qquad\qquad
+{(-1)^N\over 2}D^a\mbB_{n-1}D^a\mbgamma_{n-i-1}\Big).
}
$$
Discussions on the zero modes are given in the final section.

\chapter{The $N$-super $w_n$\ string as a broken phase of the
$(N+1)$-super
$w_n$\ string}

We consider a generalization of the superstring
hierarchy\refmark{\bop} in this
section. The $N$-super $w_n$\ string can be derived as a broken phase
of the
$(N+1)$-super $w_n$\ string. We construct a special realization of
the
$(N+1)$-super $w_n$\ algebra by adding NG superfields to generators
of the
$N$-super $w_n$\ algebra. It is shown that the cohomology of this
$(N+1)$-$w_n$\ string coincides with the one of the $N$-super $w_n$\
string by
finding a similarity transformation. Since $N+1=1,2$, and $\ge 3$\
are
different in the critical central charge \critical , we consider
these cases
separately.

\section{$N=0$\ as $N=1$}

The $N=1$\ super $w_n$\ algebra is realized by adding fermionic NG
fields
$({\eta}_i(z),{\xi}_i(z))\ (i=0,\cdots ,n-2)$\ to the generators of
the $w_n$\
algebra $W^{(m)}_i\ (i=0,\cdots,n-2)$\ as\foot{
The realization (4.1) is different from the one given in Ref.
[\kst].}
$$
\eqalignno{
W_i=&W^{(m)}_i+W_i(\eta\xi)
+\delta_{i,0}{n(n-1)\over 4}\partial^2({\xi}_0\partial{\xi}_0),\cr
V_i=&\eta_i+\sum_{j=0}^{n-i-2}\xi_jW^{(m)}_{i+j}
+\sum_{\{i;j,k\}}(i+k+1)\xi_j\partial (\xi_k\eta_{i+j+k})
+\delta_{i,0}{(2n+1)(n-1)\over 2}\partial^2\xi_0,\cr
&&\eqname\none\cr
}
$$
where
$$
W_i(\eta\xi)=\sum_{j=0}^{n-i-2}\Big(-(i+j+{3\over
2}){\eta}_{i+j}\partial{\xi}_j
-(j+{1\over 2})\partial{\eta}_{i+j}{\xi}_j\Big),
$$
and we use an abbreviated notation for multiple summations
$$
\sum_{\{i;j,k,\cdots\}}=\sum_{j=0}^{n-i-2}\
\sum_{k=0}^{n-i-j-2}\cdots .
$$

It is useful to separate the similarity transformation to three
parts\refmark{\bop}:
$$
\eqalignno{
R_1=&\oint {dz\over 2\pi i}\sum_{\{0;i,j\}}
B_{i+j}\gamma_i\xi_j,\cr
R_2=&-\oint {dz\over 2\pi i}\sum_{\{0;i,j,k\}}
(j+{1\over 2})\beta_{i+j+k}\gamma_i\xi_j\partial\xi_k,\cr
R_3=&\oint {dz\over 2\pi i}\sum_{\{0;i,j\}}
C_i\big((i+j+{3\over 2})\beta_{i+j}\partial\xi_j+(j+{1\over
2})\partial\beta_{i+j}\xi_j\big).
}
$$
These similarity transformations map the BRST charge $Q_{BRST}(1,n)$\
to the
sum of the BRST charge $Q_{BRST}(0,n)$\ and a topological charge
$Q_t$\ as
$$
e^{R_1}Q_{BRST}(1,n)e^{-R_1}=Q_1,\qquad
e^{R_2}Q_1e^{-R_2}=Q_2,\qquad
e^{R_3}Q_2e^{-R_3}=Q_{BRST}(0,n)+Q_t,
$$
where
$$
\eqalignno{
&Q_1=\oint {dz\over 2\pi
i}\sum_{i=0}^{n-2}\Big(C_i\big(W_i^{(m)}+{1\over
2}W_i(BC)+W_i(\beta\gamma)+W_i(\eta\xi)
+\delta_{i,0}{n(n-1)\over
4}\partial^2({\xi}_0\partial{\xi}_0)\big)\cr
&\qquad\qquad
+\gamma_i\big(\eta_i
-\sum_{\{i;j,k\}}(j+{1\over
2})\eta_{i+j+k}\xi_j\partial\xi_k-\delta_{i,0}{n(n-1)\over
4}\xi_0\partial^2(\xi_0\partial\xi_0)\big)\cr
&\qquad\qquad\qquad
-\sum_{j=0}^{n-i-2}\gamma_i\xi_jW_{i+j}(\beta\gamma)\Big),\cr
&Q_2=\oint {dz\over 2\pi
i}\sum_{i=0}^{n-2}\Big(C_i\big(W_i^{(m)}+{1\over
2}W_i(BC)+W_i(\beta\gamma)+W_i(\eta\xi)\big)+\gamma_i\eta_i\Big),\cr
&Q_{BRST}(0,n)=\oint {dz\over 2\pi
i}\sum_{i=0}^{n-2}C_i\big(W_i^{(m)}+{1\over
2}W_i(BC)\big),\qquad\qquad
Q_t=\oint {dz\over 2\pi i} \sum_{i=0}^{n-2}\gamma_i\eta_i.\cr
}
$$
These results reproduce those obtained in Ref. [\bo] for $N$-super
$w_2$\
string $(N,2)$.

\section{$N=1$\ as $N=2$}

Using NG $(N=1)$ superfields $({\mbeta}_i(Z),{\mbxi}_i(Z))\
(i=0,\cdots ,n-2)$\
and the generators of the $N=1$\ super $w_n$\ algebra $\mbW^{(m)}_i\
(i=0,\cdots,n-2)$, the $N=2$\ super $w_n$\ algebra is realized by the
following
infinite series\refmark{\bo}
$$
\eqalignno{
\mbW_i=&\mbW^{(m)}_i+\mbW_i(\mbeta\mbxi)
+\delta_{i,0}(n-1)\sum_{k=1}^{\infty}
\Big({-1\over 4}\Big)^k\partial\big(\mbxi_0\partial
D\mbxi_0(D\mbxi_0)^{2k-2}\big),\cr
\mbV_i=&\mbeta_i-\sum_{j=0}^{n-i-2}\mbxi_j\mbW^{(m)}_{i+j}\cr
&+\sum_{\{i;j,k\}}
\Big((i+k+{1\over 2})\mbxi_j\partial\big(\mbxi_k\mbeta_{i+j+k}\big)
+{1\over 2}D\mbeta_{i+j+k}\mbxi_jD\mbxi_k
+{1\over 4}\mbeta_{i+j+k}D\mbxi_jD\mbxi_k\Big)\cr
&\qquad\qquad\qquad\qquad
+\delta_{i,0}(n-1)\Big(\partial D\mbxi_0
+\sum_{k=1}^{\infty}\Big({-1\over 4}\Big)^kD\big(\mbxi_0\partial
D\mbxi_0(D\mbxi_0)^{2k-1}\big)\Big),
}
$$
where
$$
\mbW_i(\mbeta\mbxi)=\sum_{j=0}^{n-i-2}
\Big(-(i+j+1)\mbeta_{i+j}\partial\mbxi_j
-(j+{1\over 2})\partial\mbeta_{i+j}\mbxi_j+{1\over
2}D\mbeta_{i+j}D\mbxi_j\Big).
$$
and $\mbeta_i$\ are bosonic with dimensions $i+3/2$\ and $\mbxi_i$\
are
fermionic with dimensions $-i-1$.

The similarity transformation is now
$$
\eqalignno{
R_1=&-\oint dZ \sum_{\{0;i,j\}}
\mbB_{i+j}\mbgamma_i\mbxi_j,\cr
R_2=&\oint dZ\sum_{\{0;i,j,k\}}\Big[
\mbgamma_i\Big({1\over 2}(i+k+{1\over
2})\mbbeta_{i+j+k}\mbxi_j\partial\mbxi_k
-{1\over 4}D\mbbeta_{i+j+k}\mbxi_jD\mbxi_k\cr
&\qquad\qquad\qquad\qquad\qquad
+{1\over 8}\mbbeta_{i+j+k}D\mbxi_jD\mbxi_k
-{1\over 4}(j-k)\partial\mbbeta_{i+j+k}\mbxi_j\mbxi_k\Big)\cr
&\qquad\qquad\qquad\qquad
+\mbeta_{i+j+k}\big(-{1\over 4}(j-k)\partial\mbxi_i\mbxi_j\mbxi_k
-{1\over 8}\mbxi_iD\mbxi_jD\mbxi_k\big)\Big],\cr
R_3=&-\oint dZ\sum_{\{0;i,j\}}\mbC_i\Big(
(i+j+1)\mbbeta_{i+j}\partial\mbxi_j+(j+{1\over
2})\partial\mbbeta_{i+j}\mbxi_j
+{1\over 2}D\mbbeta_{i+j}D\mbxi_j\Big),
}
$$
which transform the BRST charge $Q_{BRST}(2,n)$ as
$$
\eqalignno{
&e^{R_1}Q_{BRST}(2,n)e^{-R_1}=Q_1,\qquad
e^{R_2}Q_1e^{-R_2}=Q_2,\qquad
e^{R_3}Q_2e^{-R_3}=Q_{BRST}(1,n)+Q_t,\cr
&Q_1=\widetilde Q_1+\sum_{k=1}^{\infty}O_k,\cr
&\widetilde Q_1=\oint
dZ\Bigg(\sum_{i=0}^{n-2}\Big[\mbC_i\big(\mbW^{(m)}_i
+{1\over 2}\mbW_i(\mbB\mbC)+\mbW_i(\mbbeta\mbgamma)
+\mbW_i(\mbeta\mbxi)\big)\cr
&\qquad\qquad\qquad\qquad
+\mbgamma_i\Big(\mbeta_i
+\sum_{\{i;j,k\}}\big(
-(j+{1\over 2})\mbeta_{i+j+k}\mbxi_j\partial\mbxi_k
+{1\over 4}\mbeta_{i+j+k}D\mbxi_jD\mbxi_k\big)\Big)\Big]\cr
&\qquad\qquad\qquad
+\sum_{\{0;i,j\}}\mbgamma_i\mbxi_j\mbW_{i+j}(\mbbeta\mbgamma)
-{(n-1)\over 4}\mbC_0\partial(\mbxi_0\partial D\mbxi_0)\Bigg),\cr
&O_k=(n-1){(-1)^k\over 4}\Big(
\mbgamma_0\mbxi_0\partial (\mbxi_0\partial D\mbxi_0(D\mbxi_0)^{2k-2})
+\mbgamma_0D(\mbxi_0\partial D\mbxi_0(D\mbxi_0)^{2k-1})
\cr&\qquad\qquad\qquad\qquad\qquad\qquad
-\mbC_0\partial (\mbxi_0\partial D\mbxi_0(D\mbxi_0)^{2k})\Big),\cr
&Q_2=\oint dZ\sum_{i=0}^{n-2}\Big[\mbC_i\big(\mbW^{(m)}_i
+{1\over 2}\mbW_i(\mbB\mbC)+\mbW_i(\mbbeta\mbgamma)
+\mbW_i(\mbeta\mbxi)
+\mbgamma_i\mbeta_i\Big],\cr
&Q_{BRST}(1,n)=\oint dZ \sum_{i=0}^{n-2}\mbC_i\big(
\mbW^{(m)}_i+{1\over 2}\mbW_i(\mbB\mbC)\big),\qquad\qquad
Q_t=\oint dZ\sum_{i=0}^{n-2}\mbgamma_i\mbeta_i.
}
$$
Here we should note the similarity transformation generated by $R_2$.
It is
convenient to consider the inverse relation since similarity
transformation is
invertible:
$$
\eqalignno{
e^{-R_2}Q_2e^{R_2}&=
{1\over k!}\sum_{k=0}^{\infty}\underbrace{[-R_2,[\cdots
,[-R_2}_{k},Q_2]\cdots]\cr
&=Q_1.
&\eqname\rtwo\cr
}
$$
Except for this transformation \rtwo, the right hand side (r.h.s.) of
similar
relations is terminated at finite order for all the other similarity
transformations. Thus all the other equations in this paper can be
confirmed by
brute force calculation. For this $R_2$\ transformation, however,
r.h.s. is not
terminated but it maps the finite series $Q_2$\ to the infinite
series $Q_1$.
This can be shown by using induction as follows. First we can show
$$
\eqalignno{
Q_2+[-R_2,Q_2]=&\widetilde Q_1,\cr
{1\over 2}[-R_2,[-R_2,Q_2]]=&{1\over 2}[-R_2,\widetilde Q_1-Q_2]\cr
=&O_1.
}
$$
This is the starting point of the induction. Then suppose
$$
{1\over
(k+1)!}\underbrace{[-R_2,[\cdots,[-R_2}_{k+1},Q_2]\cdots]=O_k.
$$
It can be shown
$$
\eqalignno{
{1\over (k+2)!}\underbrace{[-R_2,[\cdots,[-R_2}_{k+2},Q_2]\cdots]=&
{1\over k+2}[-R_2,O_k]\cr
=&O_{k+1}.
}
$$
Therefore the induction is completed.

\section{$N$\ as $N+1$\ for $N\ge 2$}

For $N\ge 2$, all the structures are common since the critical
central charges
for these cases are zero \critical. We introduce NG $N$-superfields
$(\mbeta_i,\mbxi_i)$. While $\mbxi_i$\ are always fermion with
dimension
$-i-1/2$, $\mbeta_i$\ are fermion (boson) for $N=$\ even (odd) with
dimension
$i+3/2-N/2$. By means of these NG fields and the generators of the
$N$-super
$w_n$\ algebra, $\mbW_i^{(m)}\ (i=0,\cdots ,n-2)$, we can construct
generators
of the $(N+1)$-super $w_n$\ algebra as follows.
$$
\eqalignno{
\mbW_i=&\mbW^{(m)}_i+\mbW_i(\mbeta\mbxi)\cr
=&\mbW^{(m)}_i
+\sum_{j=0}^{n-i-2}\Big(
-(i+j+{3\over 2}-{N\over 2})\mbeta_{i+j}\partial\mbxi_j
-(j+{1\over 2})\partial\mbeta_{i+j}\mbxi_j
-{(-1)^N\over 2}D^a\mbeta_{i+j}D^a\mbxi_j\Big),\cr
\mbV_i=&\mbeta_i-\sum_{j=0}^{n-i-2}\mbxi_j\mbW^{(m)}_{i+j}\cr
&+(-1)^N\sum_{\{i;j,k\}}
\big(-(i+k+1-{N\over 2})\mbxi_j\partial (\mbxi_k\mbeta_{i+j+k})
+{(-1)^N\over 2}D^a\mbeta_{i+j+k}\mbxi_jD^a\mbxi_k\cr
&\qquad\qquad\qquad\qquad
-{1\over 4}\mbeta_{i+j+k}D^a\mbxi_jD^a\mbxi_k\big).
}
$$
The similarity transformation is given by three succeeded
transformations:
$$
\eqalignno{
R_1=&-\oint dZ \sum_{\{0;i,j\}}
\mbB_{i+j}\mbgamma_i\mbxi_j,\cr
R_2=&\oint dZ \sum_{\{0;i,j,k\}}\Big[
\mbgamma_i\Big({1\over 2}(i+k+1-{N\over 2})
\mbbeta_{i+j+k}\mbxi_j\partial\mbxi_k
+{(-1)^N\over 4}D^a\mbbeta_{i+j+k}\mbxi_jD^a\mbxi_k\cr
&\qquad\qquad\qquad\qquad
+{1\over 8}\mbbeta_{i+j+k}D^a\mbxi_jD^a\mbxi_k
-{1\over 4}(j-k)\partial\mbbeta_{i+j+k}\mbxi_j\mbxi_k\Big)\cr
&\qquad\qquad
+(-1)^N\mbeta_{i+j+k}\Big({1\over
4}(j-k)\partial\mbxi_i\mbxi_j\mbxi_k
+{1\over 8}\mbxi_iD^a\mbxi_jD^a\mbxi_k\Big)\Big],\cr
R_3=&(-1)^N\oint dZ \sum_{\{0;i,j\}}
\mbC_i\Big((i+j+{3\over 2}-{N\over 2})\mbbeta_{i+j}\partial\mbxi_j
+(j+{1\over 2})\partial\mbbeta_{i+j}\mbxi_j\cr
&\qquad\qquad\qquad
-{(-1)^N\over 2}D^a\mbbeta_{i+j}D^a\mbxi_j\Big),
}
$$
which transform the BRST charge $Q_{BRST}(N+1,n)$ as
$$
\eqalignno{
&e^{R_1}Q_{BRST}(N+1,n)e^{-R_1}=Q_1,\qquad
e^{R_2}Q_1e^{-R_2}=Q_2,\qquad
e^{R_3}Q_2e^{-R_3}=Q_{BRST}(1,n)+Q_t,\cr
&Q_1=\oint dZ\Bigg(\sum_{i=0}^{n-2}\Big[
\mbC_i\Big(\mbW^{(m)}_i+{1\over 2}\mbW_i(\mbB\mbC)
+\mbW_i(\mbbeta\mbgamma)+\mbW_i(\mbeta\mbxi)\Big)\cr
&\qquad\qquad\qquad
+\mbgamma_i\Big(\mbeta_i+(-1)^N\sum_{\{i;j,k\}}
\big((j+{1\over 2})\mbeta_{i+j+k}\mbxi_j\partial\mbxi_k
-{1\over 4}\mbeta_{i+j+k}D^a\mbxi_jD^a\mbxi_k\big)\Big)\Big]\cr
&\qquad\qquad\qquad
+\sum_{\{0;i,j\}}\mbgamma_i\mbxi_j\mbW_{i+j}(\mbbeta\mbgamma)\Bigg),\
cr
&Q_2=\oint dZ \sum_{i=0}^{n-2}\Big[\mbC_i\Big(
\mbW^{(m)}_i+{1\over 2}\mbW_i(\mbB\mbC)
+\mbW_i(\mbbeta\mbgamma)+\mbW_i(\mbeta\mbxi)\Big)
+\mbgamma_i\mbeta_i\Big],\cr
&Q_{BRST}(N,n)=\oint dZ \sum_{i=0}^{n-2}\mbC_i\Big(
\mbW^{(m)}_i+{1\over 2}\mbW_i(\mbB\mbC)\Big),
\qquad\qquad
Q_t=\oint dZ \sum_{i=0}^{n-2}\mbgamma_i\mbeta_i.
}
$$
This guarantees that the cohomology of the $(N+1)$-super $w_n$\
string
coincides with the one of the $N$-super $w_n$\ string.

\chapter{Discussions}

In this paper, we have constructed a super $w$\ string hierarchy
parametrized
by two integers $(N,n)$. $N$-super $w_n$\ string can be derived as a
broken
phase of both the $N$-super $w_{n+1}$\ string and the $(N+1)$-super
$w_n$\
string. This is a generalization of two hierarchies, superstring
hierarchy\refmark{\bop} and $w$\ string hierarchy\refmark{\kst} and
including
them as subhierarchies $(N,2)$\ and $(0,n)$, respectively.

We have proved that the cohomology of the broken phase $N$-super
$w_{n+1}$\
string or $(N+1)$-super $w_n$\ string coincides with the one of the
$N$-super
$w_n$\ string. For the complete proof of equivalence, however, we
must also
show the cancellations of the zero modes, or equivalently the
coincidence of
amplitudes. This is nontrivial since the moduli spaces of two
theories are
different. Moreover, the integrand of the amplitudes are not BRST
invariant in
general but transformed to the total derivatives of the moduli. Such
a
investigation of global degrees of freedom was given in Ref. [\bv]
for $N=1$\
and 2 superstrings. It was also discussed for $w$\ string hierarchy
by giving a
generalization of the picture changing without studying the moduli
space
explicitly\refmark{\kst}. Similar consideration of the picture
changing is
possible for all the cases in the $w$\ string hierarchy. After
similarity
transformation, the Hilbert space of the broken phase string becomes
direct
product of the would-be equivalent string and a topological string.
The
difference of moduli spaces should provide the one of the topological
theory.
It is enough for the coincidence of amplitudes to show that all the
topological
string amplitudes are equal to one. This can be done, in principle,
in a
similar way to one using in Ref. [\kst]. We should give a geometrical
consideration of the moduli space in any case.

As mentioned in Introduction, the broken-phase super $w_{n+1}$\
string is
equivalent to the $w_n$\ string {\it in broken phase} since the
$w_n$\ algebra
is not a subalgebra of the $w_{n+1}$\ algebra. This is a common
structure for
the non-linear $W_n$\ algebra. If there is a hierarchy of the
non-linear $W$\
string which has a similar structure of the $w$\ string hierarchy, it
has no
further structure since subalgebra of the non-linear $W_n$\ algebra
is the only
Virasoro algebra. However, the linear $w_n$\ algebra has the other
subalgebras
which we call $w_{n/q}$\ algebra generated by $\{W_i;\ i=0\ {\rm
mod}\  q, i\le
n-2\}$. We can consider the $w_{n/q}$\ string and construct it as a
broken
phase of the $w_n$\ string. For example, the $w_4$\ algebra can be
realized by
generators of $w_{4/2}$\ algebra $\{W^{(m)}_0,W^{(m)}_2\}$\ and NG
fields
$(\beta_1,\gamma_1)$\ with dimensions $(3,-2)$\ as
$$
\eqalignno{
W_0=&W^{(m)}_0-3\beta_1\partial\gamma_1-2\partial\beta_1\gamma_1,\cr
W_1=&\beta_1-2W^{(m)}_2\partial\gamma_1-\partial
W^{(m)}_2\gamma_1,\cr
W_2=&W^{(m)}_2.
}
$$
It would be interesting to examine these symmetry breaking for
studying general
features of the $w$\ string. We hope to discuss this issue elsewhere.

There is a common feature for the generators in the broken phase
strings. The
additional (broken) generators starting from the linear term of
$\beta_i\
(\eta_i)$, which leads the corresponding $\gamma_i\ (\xi_i)$\ field
transformed
inhomogeneously by this transformation:
$\delta \gamma_i=\lambda_i+\cdots$\ ($\delta
\xi_i=\epsilon_i+\cdots$). This
means $\gamma_i\ (\xi_i)$\ is the Nambu-Goldstone boson (fermion) and
realizations constructed are two-dimensional analog of the non-linear
realization\refmark{\kun}. Since these gauge symmetries are local,
these NG
fields become unphysical\refmark{\ko,\kun}. This should be understood
in the
Lagrangian formulation by making a model which exhibits the
spontaneous
symmetry breaking. Such a investigation in the Lagrangian formulation
was given
in Ref [\b] and it was found the Lagrangian in the broken phase. It
is
interesting to find the Lagrangian in the general phase which derive
the above
broken-phase one by the spontaneous symmetry breaking.

\vskip 1cm

\centerline{\bf Acknowledgements}

The authors would like to thank Nobuyoshi Ohta for valuable
discussions. A part
of the calculations in this paper were done by using the OPE packages
developed
by Kris Thielemans and Sergey Krivonos\refmark{\t,\tk}, whose
software is
gratefully acknowledged.

\endpage

\refout
\bye